\documentclass[runningheads]{llncs}
\usepackage{color,soul}

\usepackage{enumitem}
\usepackage[dvipsnames]{xcolor}
\usepackage{graphicx}
\graphicspath{ {./images/} }
\usepackage[cmex10]{amsmath}

\usepackage{algorithm}
\usepackage[noend]{algcompatible}
\usepackage[nodisplayskipstretch]{setspace}
 \usepackage[font=scriptsize,labelfont=bf]{caption}
\begin{document}
\title{Context-Aware Trustworthy IoT Energy Services Provisioning\vspace{-10pt}}

\author{Amani Abusafia \inst{1} \orcidID{0000-0001-9159-6214}\and
Athman Bouguettaya \inst{1} \orcidID{0000-0003-1254-8092} \and
Abdallah Lakhdari \inst{1} \orcidID{0000-0001-8005-1534} \and
Sami Yangui \inst{2} \orcidID{0000-0001-9756-642X}}
\authorrunning{A. Abusafia et al.}
\institute{The University of Sydney, Sydney NSW 2000, Australia\and LAAS-CNRS, Université de Toulouse, INSA, 31400 Toulouse, France\\
\email{\{amani.abusafia, athman.bouguettaya, abdallah.lakhdari\}@sydney.edu.au}\\yangui@laas.fr\vspace{-10pt}}

%
%
%



\maketitle              
\begin{abstract}


We propose an IoT energy service provisioning framework to ensure consumers' \textit{Quality of Experience (QoE)}. A novel  \textit{context-aware trust assessment} model is proposed to evaluate the trustworthiness of providers. Our model adapts to the dynamic nature of energy service providers to maintain QoE by selecting trustworthy providers. The proposed model evaluates providers' trustworthiness in various contexts, considering their behavior and energy provisioning history. Additionally, a trust-adaptive composition technique is presented for optimal energy allocation. Experimental results demonstrate the effectiveness and efficiency of the proposed approaches.\looseness=-1


\vspace{-8pt}
\end{abstract}
\small
\vspace{-10pt}
\section{Introduction}
\vspace{-5pt}

 
\emph{Energy-as-a-Service (EaaS)} refers to the wireless delivery of energy from an energy provider (e.g., a smart shoe) to a nearby energy consumer (e.g., a smartphone)\cite{lakhdari2020composing}. 
Energy service may enable an eco-friendly self-sustained environment by exchanging \textit{spare} or \textit{harvested} energy \cite{dhungana2020peer}\cite{li2023activity}. For instance, an energy provider may offer their harvested energy to a nearby IoT device. Energy may be harvested from natural resources, e.g., physical movement \cite{li2023activity}. For example, wearing a PowerWalk harvester may produce energy from an hourly walk at a comfortable speed to charge up to four smartphones \cite{abusafia2022maximizing}.  
Moreover, energy services offer a \textit{convenient} and \textit{ubiquitous} power access for IoT users without using cords or power banks \cite{fang2018fair}. Energy services may be deployed through the newly developed ``Over-the-Air" wireless charging technologies \cite{abusafia2022services}\cite{OvertheAirCharger}. Several companies are developing charging technologies that enable IoT devices to charge wirelessly over a distance, such as Xiaomi, Energous, and Cota \cite{abusafia2022maximizing}\cite{lakhdari2020Vision}. For instance, Energous developed a device that can charge up to 3 watts of power within a 5-meter distance.  Although current technology may not provide efficient energy delivery \cite{feng2020advances}, technological advances are expected to enable devices to exchange larger amounts of energy \cite{abusafia2022services}.\looseness=-1



\begin{figure}[!t]
    \centering
     \setlength{\abovecaptionskip}{2pt}
    \setlength{\belowcaptionskip}{-20pt}
        \includegraphics[width=0.6\linewidth]{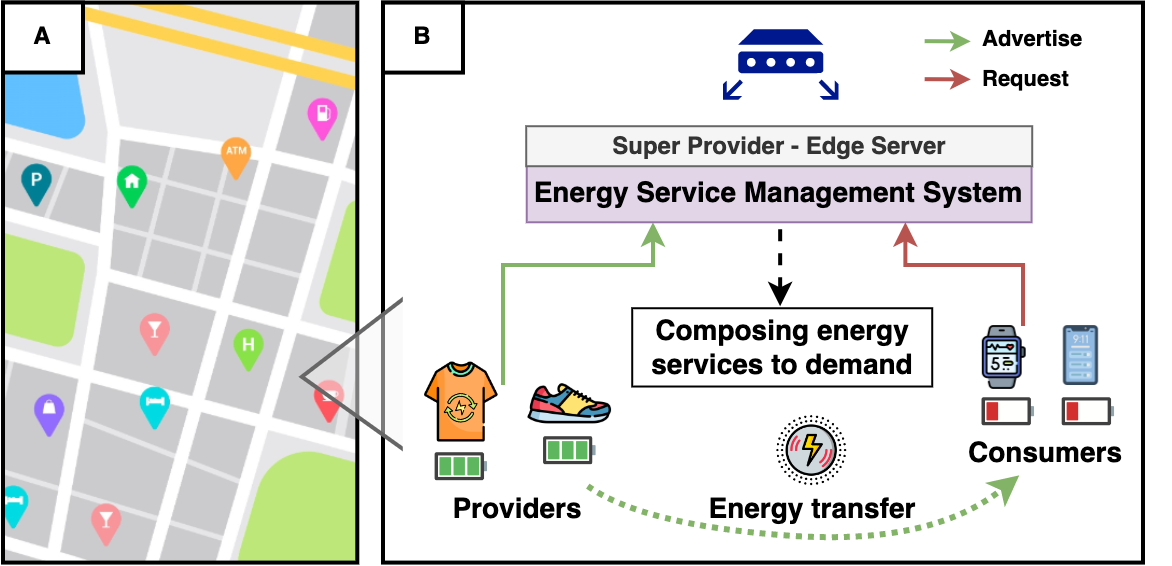}
    \caption{(A) Microcells in a smart city (B) IoT energy services environment in a microcell}
    \label{fig:scenario}
\end{figure}

We propose a dynamic energy service ecosystem that consists of energy providers and consumers in \textit{microcells} (see  Fig. \ref{fig:scenario}(A)). A microcell is any confined space where people may gather, e.g., cafes and restaurants. In this environment, IoT devices may share energy with nearby devices. The energy Service Oriented Architecture; SOA-based business model has three main actors: energy \textit{provider} \footnote{\textcolor{black}{We used interchangeably the terms energy provider and provider to refer to
the energy provider}}, energy  \textit{consumer}, and \textit{super-provider} \cite{abusafia2022services}. 
According to this business model, providers advertise their services, consumers submit their requests, and the super-provider (i.e., the microcell's owner) manages the exchange of energy services between providers and consumers in the microcell. This paper focuses on energy services sharing from the super-provider perspective.\looseness=-1

Recent research suggests that super-providers may use energy services to \textit{enhance consumers' Quality of Experience (QoE)} \cite{abusafia2022quality}\cite{abusafia2022maximizing}.  Studies show that businesses providing wireless energy services, like ``air-charge", positively affect customer experiences\footnote{air-charge.com}.
In energy services, \textit{Quality of Experience (QoE)} refer to the \textit{aggregated satisfaction} of consumers with energy services \textit{over time} \cite{abusafia2022quality}\cite{abusafia2022maximizing}. Consumers' satisfaction is measured by the fulfillment of their energy needs \cite{abusafia2022quality}\cite{abusafia2022maximizing}.  This paper focuses on energy services provisioning as a key ingredient to provide customers with the best QoE.\looseness=-1


\textcolor{black}{A key challenge in QoE-based energy service provisioning is to \textit{assess providers' commitment} to sharing energy \cite{abusafia2022services}\cite{lakhdari2020Vision}.  For example, a provider may terminate the energy transfer by leaving the transfer range. They may also stop the transfer due to excessive device usage.  Such service disruptions may reduce the expected amount of shared energy, thereby impacting the consumer's QoE. In such cases, real-time service replacement may not be guaranteed \cite{lakhdari2020Elastic}. Thus, a super-provider needs to assess providers' trustworthiness before allocating services to consumers. Hence, a \textit{trust assessment framework is necessary to evaluate the uncertainty in a provider's commitment.}}\looseness=-1 



\textit{Trust} refers to the belief that providers will adhere to agreements and maintain the quality of service as advertised \cite{lakhdari2020Vision}\cite{abusafia2022services}.  \textcolor{black}{Existing trust frameworks are hardly applicable to crowdsourced IoT energy services environments \cite{dhungana2020peer}\cite{abusafia2022services}. This is mainly due to the highly dynamic and fluctuating energy provisioning and usage behavior of IoT users \cite{lakhdari2020fluid}. Consequently, energy fluctuations directly result from the uncertainty around crowdsourced IoT energy providers meeting their energy commitments. As indicated, IoT users’ commitment may fluctuate due to the mobility and usage patterns of IoT users \cite{abusafia2022services}\cite{lakhdari2020fluid}.  For instance, an IoT user may consume their advertised service due to unexpected heavy device usage \cite{lakhdari2020Elastic}. Given IoT users' energy fluctuation and dynamic behavior, using existing trust frameworks may often lead to low trust scores for most providers.  If most providers have low trust scores, the super-provider may not find enough trustworthy services to allocate.  For example, existing trust frameworks would typically give a low trust score to a provider who consistently offers only 50 mAh, regardless of the advertised services. However, if the super-provider only needs 50mAh, assessing the provider based on this specific requirement would result in a higher trust score, making them a suitable candidate for allocation. Therefore, the same IoT energy provider’s trust will vary according to the super-providers’ requirements. Therefore, a new trust assessment framework is needed \cite{abusafia2022services}\cite{lakhdari2020Vision}. In this respect, we propose a context-aware trust assessment framework tailored specifically to energy services.}\looseness=-1

\textcolor{black}{We propose a context-aware trust assessment framework to accurately assess and effectively allocate trustworthy providers. The framework assesses providers' trustworthiness based on the super-provider context-aware constraints.} The providers' trust scores are then used to \emph{select} and \emph{compose} the best set of energy services to fulfill the super-provider requirements. Our framework enables utilizing providers that may appear untrustworthy but can make valuable contributions while maintaining commitment. In addition, we propose a heuristic-based approach that ensures a higher QoE. Our approach composes additional services, based on the trust scores of providers, as a backup in the case of service cancellation. The main contributions of this paper are:\looseness=-1
\begin{itemize}[noitemsep,nosep,leftmargin=4pt,labelsep=2pt,itemindent=2pt]
    \item A novel trust-aware framework to compose energy services and ensure QoE.
    \item \textcolor{black}{A context-aware trust assessment model to evaluate energy providers.} 
    \item A context model for defining super-provider constraints in evaluating provider trust.
     \item A heuristic-based approach to compose trustworthy energy services.
\end{itemize}

\begin{figure}[!t]
    \centering
     \setlength{\abovecaptionskip}{2pt}
    \setlength{\belowcaptionskip}{-10pt}
        \includegraphics[width=0.45\linewidth]{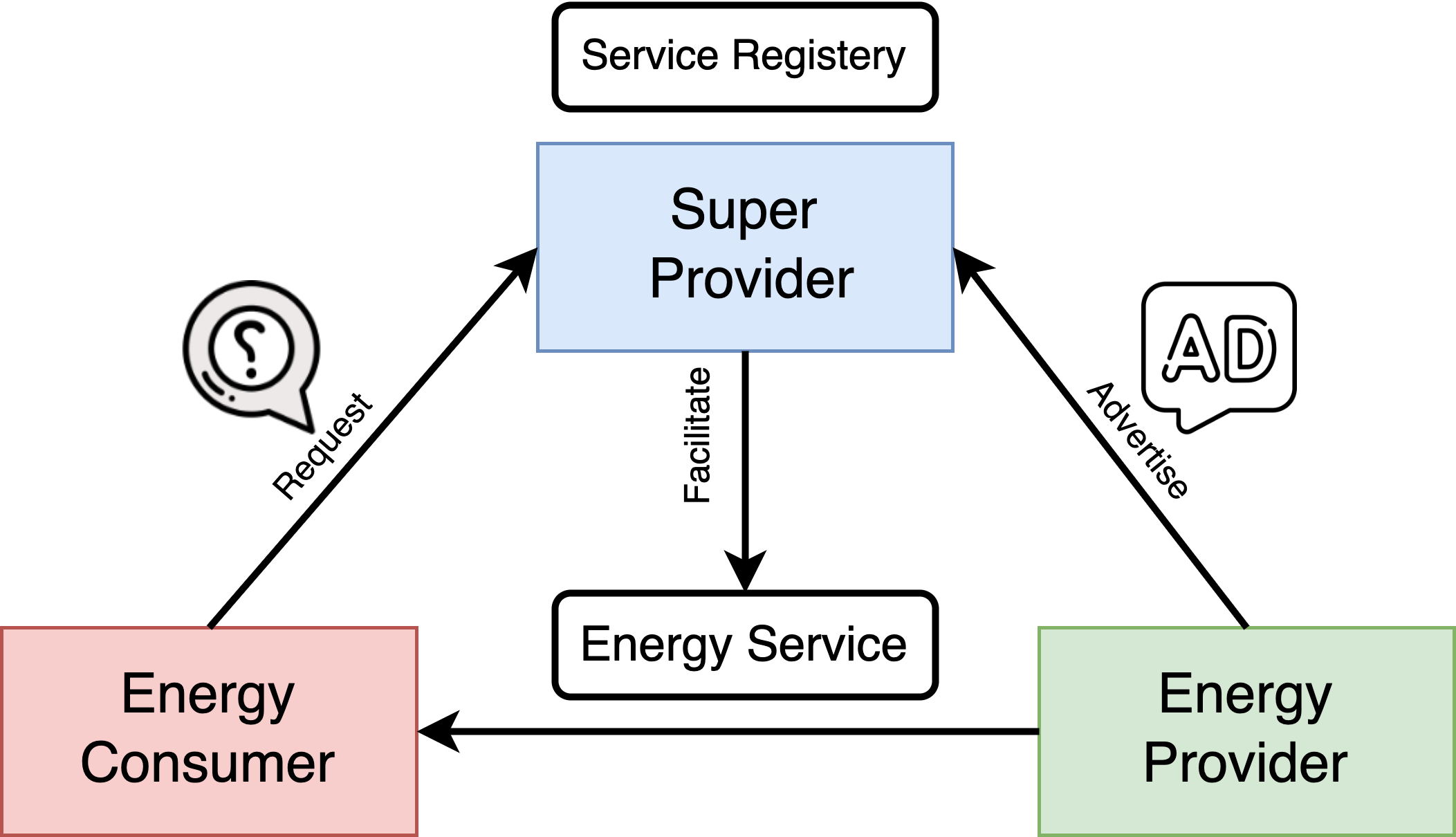}
    \caption{IoT energy services business model}
    \label{fig:Bmodel}
\end{figure}

\vspace{-15pt}
\subsubsection*{Motivating Scenario}:
We describe a scenario in a confined place (i.e., microcell) where people congregate, e.g., cafes and restaurants (see Fig. \ref{fig:scenario} (A)). Each microcell may have several IoT devices acting as energy providers or consumers (see Fig. \ref{fig:scenario} (B)). The super-provider aims to leverage the crowdsourced energy services to enhance the consumers' \textit{experience}. \textcolor{black}{We assume that all energy services and requests are sent to the \textit{edge} (e.g., a microcell router, Fig. \ref{fig:scenario}(B)) and managed by the super-provider (Fig. \ref{fig:Bmodel})}.
We assume that the {super-provider} has prior knowledge of the \textit{energy demand distribution} in the microcell over a given time window. Additionally, the super-provider adopts a QoE-based approach, such as \cite{abusafia2022maximizing}\cite{abusafia2022quality}, to select optimal providers to fulfill the microcell's energy demand. \textcolor{black}{We assume that the super-provider offers incentives to encourage energy sharing in the form of credits \cite{abusafia2022quality}\cite{abusafia2020incentive}. The credits would be used to receive more energy when the providers act as consumers in the future \cite{abusafia2022maximizing}}. However, rewards do not guarantee providers' commitment \cite{abusafia2020incentive}. \textcolor{black}{The \textit{uncertainty} in providers' behavior may adversely affect consumers' QoE if a service fails and no other nearby services are available \cite{lakhdari2020Elastic}. Thus, a trust framework is required to evaluate providers' commitment. While there are several trust frameworks for IoT services, they are not applicable to energy services as they do not accommodate the nature of energy services. An inaccurate trust assessment may lead to misjudgment of a service. This could lead to either not using it due to a low score or using an untrustworthy service that disappoints consumers.  Given energy scarcity,  it is paramount to assess trust in order to efficiently utilize all available services correctly. Furthermore, super-providers typically may vary in their service expectations and requirements. These expectations may result in different trustworthiness scores for the same provider. Hence, it is challenging to find the right trust assessment that meets super-provider expectations.}\looseness=-1

\begin{figure*}[!t]
    \centering
       \setlength{\abovecaptionskip}{3pt}
    \setlength{\belowcaptionskip}{-5pt}
    \includegraphics[width=0.9\linewidth]{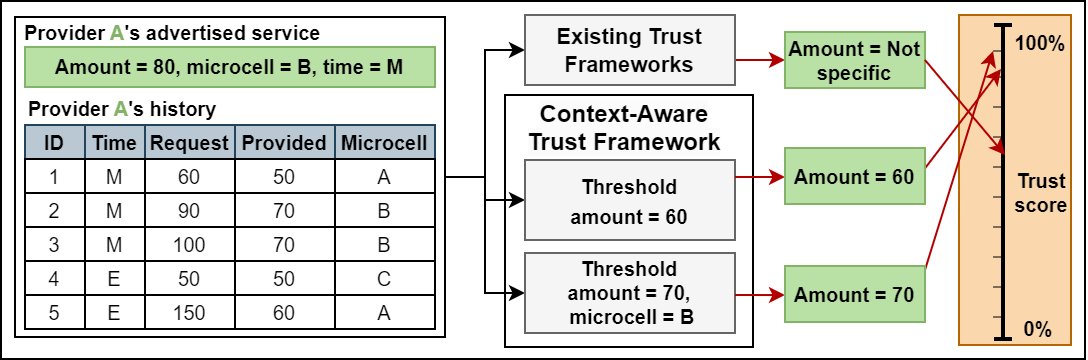}
    \caption{Example of how different trust assessments yield different trust scores.
}
    \label{fig:fullscenario}
    \vspace{-12pt}
\end{figure*}

Fig. \ref{fig:fullscenario} shows a provider's advertised service and history. Let us assume we limit the trust metrics to the provider's commitment to fulfill an energy request. Using a traditional trust framework where a provider is assessed based on all its available history, the provider trust score will be 67\% without estimating the provided amount. \textcolor{black}{This score is computed based on the ratio of the total provided energy to the total requested energy.}   However, using a more context-aware trust framework, where a super-provider defines their minimum requirements as a threshold,  may result in a different score. For example, if a super-provider has a 60 mAh threshold, they will look at the provider's history and how often they fulfill 60 mAh requests. Based on this, they rated the provider's service to 60 mAh with a 93\% trust score. In another example, if the threshold is set to evaluate the rate of providing 70mAh services in microcell B, the trust framework will consider the provider's history within microcell B to understand their performance in that area. As a result, the super-provider will rate the provider's services as 70mAh with a 100\% trust score.\looseness=-1

Allocating the \textit{limited} available energy amid the \textit{uncertainty} of providers' commitment presents a critical challenge for the efficient and QoE-aware provisioning of IoT energy services \cite{abusafia2022services}. We propose a context-aware trust-assessment framework for energy services. Our framework assesses provider trust based on the expectations and requirements of super-providers, which may lead to varying trust scores for the same service. The framework composes the most trustworthy energy services, ensuring consumer QoE and leveraging super-providers' expectations to assess and adhere to their standards. Moreover, our composition approach accounts for the natural untrustworthiness in the energy crowdsourcing environment by selecting additional services as a backup. The selection of these services relies on the trustworthiness of the available services\looseness=-1.
\vspace{-10pt}
\section{System Model}
\vspace{-10pt}
We focus on assessing the trust of energy services in microcells during a single time slot $t$. We use the below definitions to formalize the problem.
\vspace{-10pt}
\subsection{Energy Service Model}
\vspace{-5pt}
\begin{definition} \label{ESdef}
\textit{\textbf{Energy Service (S)}}. We adopt the definition of
EaaS introduced in \cite{lakhdari2020composing}. An energy service \textit{S is defined as a tuple of $<s_{id}, p_{id}, F, Q>$,  where:}
 \begin{itemize}[ noitemsep,nosep,leftmargin=8pt,labelsep=2pt,itemindent=2pt]
     \item $s_{id}$ is a unique service identifier, 
     \item $p_{id}$ is a unique provider identifier,
     \item $F$ is the function of delivering wireless energy,
     \item $Q$ is a tuple of $< q_1, q_2, ..., q_n>$, where each $q_i$ denotes the Quality of Service (QoS), e.g., energy amount and geographical location.
 \end{itemize}
\end{definition}
\vspace{-15pt}
\begin{definition}\label{ESQoSdef}
\textit{\textbf{Energy Quality of Service (QoS).}} \textit{QoS attributes enable users to differentiate between energy services \cite{lakhdari2020Vision}. We extend the definition of QoS attributes from \cite{lakhdari2020composing}. QoS is presented as a tuple of $<a, l, d, b>$ \cite{lakhdari2020composing}, where:}
     \begin{itemize}[noitemsep,nosep,leftmargin=8pt,labelsep=2pt,itemindent=2pt]
            \item $a$ is the amount of energy offered by the provider,
            \item $l$ is the geographical location of the provider,
            \item \textcolor{black}{$d$ is a tuple  of $<st, et>$ that represents the service's start and end time,}
            \item $tr$ is the provider's trust score to offer their service,
        \end{itemize}
\end{definition}
\vspace{-15pt}
\begin{definition}\textit{\textbf{Energy Service Provider (P).}} \textit{P is an IoT device with spare energy to be shared as a service. P is defined as a tuple of $<p_{id}, \mathcal{H}>$,  where:}
 \begin{itemize}[ noitemsep,nosep,leftmargin=8pt,labelsep=2pt,itemindent=2pt]
     \item $p_{id}$ is a unique provider identifier,
     \item $\mathcal{H}$ is the set of historical records of the provider's previous provisioning.  We assume that all providers have a history, and dealing with newcomers is not the focus of this paper. We also assume that the edge will retrieve the history of providers from the cloud (see  Fig. \ref{fig:scenario}(B)). $\mathcal{H}$ is a tuple of $< h_1, h_2, ..., h_m>$, where each $h_i$ represent the record of previous energy transfer as $<s,de,m,t>$ where:
     \begin{itemize}[ noitemsep,nosep,leftmargin=8pt,labelsep=2pt,itemindent=2pt]
        \item \textcolor{black}{$s$} is the advertised energy service,
        \item $de$ is the amount of delivered energy,
        \item $m$ is the microcell where the energy sharing occurred,
        \item $t$ is the time \textcolor{black}{interval} $<st, et>$ of the energy sharing.
    \end{itemize}
 \end{itemize}
\end{definition}
\vspace{-15pt}

 \vspace{-10pt}
\subsection{Super-Provider Preferences}
\vspace{-5pt}
\label{SPP}
\begin{definition}
\textbf{\textit{Energy Demand Distribution (ED).}} The super-provider uses the traffic history of the microcell to define the Energy Demand distribution ($ED$). \end{definition} \vspace{-5pt}
We adopt the definition of $ED$ from \cite{abusafia2022maximizing}. \textit{$ED$ is the predicted consumers' energy demand distribution in the service time window $W$.} Therefore, the demand for a single time slot is defined as a tuple $<d,t>$ where $d $ is the aggregation of the predicted energy requests at time interval $t$. The prediction of energy requests may be obtained using prediction techniques applied to the historical records $\mathcal{H}$ for all IoT users who visited the microcell at time interval $t$ \cite{mach2017mobile}\cite{abrishami2018using}. Energy requests may be aggregated according to their spatio-temporal features \cite{lakhdari2020Elastic}. Given a set of predicted energy requests, the super-provider aggregates the energy requests using the composition approach proposed by \cite{lakhdari2020composing}. The approach considers the time interval of each request to define a composite energy request that includes all available requests. The super-provider sums the requested energy  by all the available requests at that time interval using:\looseness=-1
\vspace{-8pt}
\begin{equation}
 \label{eq:energy_demand}
        \textcolor{black}{Ed.d(t)} = \sum_{j=0}^m f(\mathcal{H}_j,t) \vspace{-8pt}
 \end{equation}

Where $d$ is the aggregated energy demand at time slot $t$, $\mathcal{H}_j$ is the provider's $j$ history of energy requests at time $t$, $m$ is the number of consumers who have visited the microcell at $t$ in the past and $f$ is the prediction function.\looseness=-1
\vspace{-5pt}

\begin{definition}
\textbf{\textit{Quality of Experience (QoE).}} QoE measures the aggregated satisfaction of energy consumers in a microcell at time slot $t$.\end{definition} \vspace{-5pt} The consumers’ satisfaction is determined by the allocated services to the energy demand  $ED$. We adopt the definition of QoE from \cite{abusafia2022quality}, which measures QoE across time slots and adjusts it to assess a single time slot. Therefore, QoE is computed as:\looseness=-1\vspace{-10pt}
\begin{equation}
    QoE = \sum_{i=1}^n S_i.a /\textcolor{black}{Ed.d(t)}
    \vspace{-8pt}
\end{equation}

 Where $n$ is the number of selected services, $a$ is energy service $i$'s amount, 
 \textcolor{black}{${Ed.d(t)}$ is the aggregated energy demand at time slot $t$ computed using Eq. \ref{eq:energy_demand}}, $m$ is the number of aggregated requests, and $re$ is energy request size.\looseness=-1


\vspace{-15pt}
\subsection{Problem Definition}\label{prblemform}
\vspace{-5pt}
Given a super-provider in a microcell $m$ who wants to fulfill their consumer's expected energy demand $Ed.d(t)$ at a time slot $t$.  The super-provider has a set of $n$ candidate providers $\mathcal{P} =\{p_1, p_2, ..., p_n\}$ who expressed their interest in offering their service  $\mathcal{S} =\{s_1, s_2, ..., s_n\}$ at that time and location.  Each provider has a history $\mathcal{H} = < h_1, h_2, ..., h_m>$ of sharing energy.
The super-provider aims to ensure QoE by maximizing the fulfillment of energy demand. This will be achieved by allocating the most trustworthy providers. 
\textit{We reformulate the service provision problem as a time-constrained optimization problem} as follows:
\begin{itemize}[noitemsep,nosep]
    \item Maximize $QoE$ as $\sum_{i=1}^n S_i.a *\mathcal{P}_{{Trust}_i} /\textcolor{black}{Ed.d(t)}$
  \end{itemize}
Subject to:
\begin{itemize}[ noitemsep,nosep]
     \item $Ed.d(t)> 0$,
    \item $S_{i}.d \subset t$  for each $S_{i}  \in \mathcal{S}$.
\end{itemize}
Where $QoE$ is computed based on the allocated energy multiplied by the corresponding provider trust score, $\mathcal{P}_{{Trust}_i}$ is provider $i$ trust score, $S_{j}.d$ is the time interval $<st, et>$ a provider of $S_{i}$  may offer their energy, $ t$ is the duration of a time interval, and  $Ed.d$ is the aggregated energy demand at time slot $t$.\looseness=-1

\noindent We use the following assumptions to formulate the problem:
\begin{itemize}[noitemsep,nosep,leftmargin=4pt,labelsep=2pt,itemindent=2pt]
\item Providers' energy size is fixed during composition.
\item Providers and consumers are static during energy sharing.
\item The super-provider context model is given as input, and determining the constraints is out of the scope of this paper. 
\item There is no energy loss in sharing. As the technology matures, we anticipate that the devices will share more energy, and the sharing energy loss will become minimal \cite{abusafia2022maximizing}.\looseness=-1 
\item  \textcolor{black}{The super-provider uses credits as incentives for energy sharing, which providers can later redeem for more energy when they become consumers \cite{abusafia2022quality}.}

\item \textcolor{black}{All providers have a history of energy provisioning and are willing to share them}.\looseness=-1
\item The energy demand distribution  $ED$  is deterministic.

\item A secure framework has been implemented to preserve the integrity and the privacy of the  IoT devices \cite{zhang2020charging}.\looseness=-1
\end{itemize}

\vspace{-12pt}
\section{Trust Assessment Model} 
\label{trustMdl}
\vspace{-10pt}
We define the provider trust level assessment to determine the trustworthiness score ($\mathcal{P}_{Trust}$) of an energy service.  The trust level assessment considers the behavior of providers and their history in delivering energy. We compute the trustworthiness of a provider  using the following attributes:
\begin{itemize}[noitemsep,nosep,leftmargin=0pt,labelsep=1pt,itemindent=0pt, labelwidth=*]
\item \textbf{Success Rate:} \textcolor{black}{This attribute measures the reliability of a provider based on their past performance. We argue that providers with a high success rate are likely to be more reliable in the future, making this an important attribute in trust evaluation.} The \emph{success rate} ($\mathcal{SR}_P$ ) of a provider is  the ratio of \emph{completed energy services} to the total number of \emph{initiated services} by that provider. Here, completed services refer to full energy delivery as advertised, while initiated services count all, regardless of completion. We compute $\mathcal{SR}_P$ of provider $P$ as follows:\looseness=-1
\vspace{-7pt}
\begin{equation}
    \mathcal{SR}_P = \frac{|\{S \in \mathcal{E}_P \;|\; S\;is\; completed\}|}{|\mathcal{E}_P|}
    \vspace{-8pt}
\end{equation}

where $\mathcal{E}_P$ is all services initiated by provider $P$ and $|.|$ is the cardinality of the set.\looseness=-1 

\item \textbf{Delivery Size:} \textcolor{black}{This attribute gauges the consistency of a provider in terms of the quantity of energy they deliver.} The Delivery Size rate ($\mathcal{DS}_{P}$) may be calculated as the ratio of 
successful energy deliveries' total size to all attempted deliveries' total size by a provider. $\mathcal{DS}_{P}$ is computed using:

\vspace{-3pt}
\begin{equation}\label{eq:ds}
\mathcal{DS}_{P} = \frac{\sum_{i=1}^{n} h_i.de}{\sum_{i=1}^{n} S_i.a}
\vspace{-3pt}
\end{equation}
where $n$ represents the number of previous services delivered by the provider, $h_i.de$ represents the actual delivered energy retrieved from the history record $h_i$, and  $S_i.a$ represents the advertised amount of the $i$-th energy service.

\item \textbf{Timeliness Score:} \textcolor{black}{This attribute measures a provider's adherence to the service schedule.  Disconnections in the transfer process may sometimes occur due to the provider's indoor movement, leading to energy transfer delays \cite{lakhdari2020fluid}. These delays, increasing consumer wait times, can negatively impact their QoE.} Hence, the Timeliness score $\mathcal{TL}_{P}$ is crucial in the trust model. $\mathcal{TL}_{P}$ is calculated using the following formula:
\vspace{-3pt}
\begin{equation}
    \mathcal{TL}_{P}
    \begin{cases}
        1 & \text{if $ \sum_{i=1}^n {(h_i.t.et-S_i.et)} <= 0$} \\
        \frac{1}{\sum_{i=1}^n {(h_i.t.et-S_i.et)}/n} & \text{otherwise}
    \end{cases}
    \vspace{-3pt}
\end{equation}

where $n$ is the provider's number of previously delivered energy services,  $h_i.t.et$ represents the actual end time of delivering a service $i$ retrieved from the provider's history record $h_i$, and $S_i.et$ represents the advertised end time of the $i$-th energy service.\looseness=-1

\item \textbf{Impact Score:} \textcolor{black}{A provider's service cancellation may affect consumers' QoE. As a single service may be fulfilling multiple requests \cite{lakhdari2020Elastic}, the impact of canceling service on the consumers may be used to evaluate the providers' trustworthiness.} We compute the service failure's impact $\mathcal{F}_{S}$ based on the number of affected consumers.  We compute the failure impact $\mathcal{F}_{S}$ of a \textit{canceled} energy service as follows:
\vspace{-6pt}
        \begin{equation}
        \label{Feffect}
            \mathcal{F}_{S} = \frac{|\{C \in c\; |\; C\;is\; receiving\; from\; S}{|C|}
            \vspace{-10pt}
        \end{equation}
 where $C$ is the set of all consumers in the microcell within the duration of the service [$S_{st} - S_{et}$] and $|.|$ is the cardinality of a set. 

 We define the provider impact as the cumulative effect of all their services on consumers. Typically, a trustworthy provider will have minimal impact on consumers when a failure occurs. In other words, the provider's impact is calculated as the complementary value to the failure impact. Therefore, We compute the provider impact score $\mathcal{I}_{P}$ as follows:
 \vspace{-6pt}
\begin{equation}
      \mathcal{I}_{P} = \left(\sum_{i=1}^{n}(1-\mathcal{F}_{{S}_{i}})\right)/n
      \vspace{-6pt}
\end{equation}
Where $n$ represents the number of previous energy services delivered by the provider.

\item \textbf {Mobility Pattern:} \textcolor{black}{A provider's mobility pattern may influence their energy service provision. For example, if a provider's mobility pattern shows their staying time at a microcell is 5 minutes, and they advertised a service that requires 30 minutes, this might indicate the probability of not committing to the request.} Therefore, we consider the staying duration pattern as an attribute in the trust assessment. We use the history of the provider's previous services to compute their duration patterns \cite{gonzalez2008understanding}. The number of previous services should be greater than a predefined threshold $\alpha$ to be considered. We compute the staying duration pattern $\mathcal{SD}_{P}$ as:
\vspace{-8pt}
\begin{equation}
    \mathcal{SD}_{P} = \left(\sum_{i=1}^{k} (S_{i}.{et} - S_{i}.{st})\right)/k \;|\; k > \alpha
    \vspace{-8pt}
\end{equation}
where $S_{i}.{st}$ is the provider's previous service $i$ start time,  $S_{i}.{et}$ is the provider's previous service $i$ end time, and $k$ is the number of previous services. We compute the provider Duration trust factor $\mathcal{D}_{P}$ using the staying duration pattern $\mathcal{SD}_{P}$ as follows:
\vspace{-6pt}
\begin{equation} \label{eq:du}
\mathcal{D}_{P} =
\begin{cases}
    1 & \text{if $(S.{et} - S.{st}) <= \mathcal{SD}_{P}$} \\
    \frac{SD_C}{(S.{et} - S.{st})} & \text{otherwise}
    \end{cases}
    \vspace{-6pt}
\end{equation}
where $S.{st}$ is the provider's current service start time,  $S.{et}$ is the provider's current service end time.
\end{itemize}

\vspace{-15pt}
\subsection*{Providers Trustworthiness}
\label{PTrustSec}
\vspace{-10pt}
\textcolor{black}{The provider trust score ($\mathcal{P}_{Trust}$) combines all the aforementioned trust attributes to offer a holistic assessment of a provider's trustworthiness. While each attribute reflects different aspects of a provider's service delivery, their significance varies depending on the super-provider-specific requirements. For example, a super-provider may ignore the timeliness score if they have customers who stay for a long time. Hence, we added a component in the trust-assessment constraints for the super-provider to determine attribute weights.} $\mathcal{P}_{Trust}$ is calculated using the following formula:\vspace{-11pt}
\begin{equation}
\label{eq:PTrust}
  \mathcal{P}_{Trust} = \sum_{J \in l}^{}w_{J}\times {\mathcal{L}_P} 
  \vspace{-11pt}
\end{equation}
 where $\mathcal{L} =\{{SR, TL, DS, I, D}\}$ and $w$ represents the weights of each trust attribute based on its importance, with the constraint that $\sum w_i= 1$  to normalize the impact of each factor on the overall trustworthiness score. Multiple methods exist for calculating the weight of each attribute, depending on the preference of the super-provider~\cite{bahutair2021multi}.\looseness=-1 
\vspace{-10pt}

\begin{figure}[!t]
    \centering
     \setlength{\abovecaptionskip}{2pt}
     \setlength{\belowcaptionskip}{0pt}
   \includegraphics[width=\linewidth]{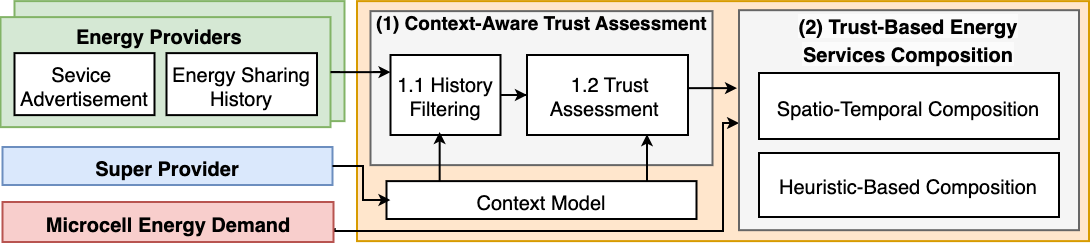}
    \caption{Trust-based energy service composition framework }
    \label{fig:frameWork}
  \end{figure}





\vspace{-8pt}
\section{Energy Service Composition Framework}
\vspace{-10pt}
\textcolor{black}{We introduce our energy service composition framework for allocating trustworthy energy services to provide consumers with the best QoE  (see  Fig. \ref{fig:frameWork}). The framework ensures QoE by allocating trustworthy services based on the super-provider context. The framework is divided into two phases: (1) context-aware trust assessment and  (2) trust-based energy service composition. The general steps of the framework are present in Algorthim \ref{alg:TBC}. }
In what follows, we discuss each phase in detail:\looseness=-1

\begin{algorithm}[t!]
\footnotesize
\setlength{\textfloatsep}{-50pt}
\setlength{\floatsep}{0pt}
  \textcolor{black}{  \caption{Trust-Based Energy Services Composition}
    \label{alg:TBC}
    \begin{algorithmic}[1]
        \renewcommand{\algorithmicrequire}{\textbf{Input:}}
        \renewcommand{\algorithmicensure}{\textbf{Output:}}
        \REQUIRE $Energy\; Providers(P), Energy\;Demand(ED), Context\; Model(CM)$
        \ENSURE $ES_{comp}, QoE$ 
        \Statex 
        \textbf{Phase 1: Context-Aware Trust Assessment}\label{Aendph1}
        \FOR{$p\; in\; P$}\label{Astartph1} 
            \IF{$ [p.st,p.et] \subseteq Ed.t$}
                \State $ p.h = History\_Filtering(p.\mathcal{H},CM)$
                \State $ \mathcal{P}_{Trust} = Trust\_Assessment(p.h,CM)$
                \State $ newE = p.s.a *  \mathcal{P}_{Trust}$
                \State $SelectedP.add(p,\mathcal{P}_{Trust},newE)$
             \ENDIF
        \ENDFOR
        \Statex 
        \textbf{Phase 2: Composition of Energy Services}
        \State $ES_{comp} = Energy\_Allocation(SelectedP,ED)$  
        \State $QoE = QoE\_Assessment(ES_{comp},ED)$    \State \textbf{return } $ES_{comp}, QoE$ \label{Aendph2} 
  \end{algorithmic}}
\end{algorithm}
\setlength{\textfloatsep}{3pt}
\vspace{-8pt}
\subsection{Context-Aware Trust Assessment}\label{TrustAssesment}
\vspace{-8pt}
In this phase, the proposed framework evaluates the historical performance of each provider to assess their trustworthiness. This framework uses our proposed trust model to evaluate a provider's trust.  It further incorporates a context model, which represents the constraints set by the super-provider regarding the attributes and data used for the trust assessment. The context model consists of two main components: (1) history constraints and (2) trust assessment constraints. This model is then used in the trust assessment phase (see Fig. \ref{fig:frameWork}). In the following subsections, we will present the context model and discuss each step of the framework in detail.\looseness=-1

\vspace{-15pt}
\subsubsection{Context Model}
As mentioned, the context model establishes the constraints set by the super-provider concerning the attributes and data used in evaluating providers' trustworthiness. The context model consists of two sets of constraints. The first set pertains to the provider's history, addressing aspects such as considering their entire history or focusing solely on their behavior within a specific microcell. The second set defines the super-provider's constraints on trust assessment attributes, which may include using all trust model attributes or only a subset, adjusting the weights of trust model attributes, or evaluating particular attributes. \textcolor{black}{While we provide a comprehensive model to represent the possible constraints of super-providers, we assume that the selection of which constraints to apply and the values of the variables in these constraints will be provided as input to our framework. These values will be determined based on the business context and requirements of the super-provider.} In what follows, we discuss each constraint.


\begin{itemize}[noitemsep,nosep,leftmargin=8pt,labelsep=1pt,itemindent=0pt, labelwidth=*]
    \item \textbf{History Constraints:} As previously mentioned, trust assessment can employ the provider's entire history or part of it, depending on the super-provider history constraints (see  Fig. \ref{fig:TrustAssesment}). For instance, a super-provider may assess providers' trust based on their history within a specific microcell. Another example would be evaluating the provider's energy-sharing history for larger requests, such as those exceeding 100 units. Therefore, we formulate the history constraints as a constraint satisfaction problem (CSP) \cite{kumar1992algorithms}. Consequently, the super-provider history constraints are represented as a triple $<X, D, C>$, where X is a set of variables, D is a set of corresponding domains of values, and C is the super-provider's constraint set. The formulation of the super-provider history constraints is as follows:
    \vspace{-3pt}
    \begin{equation}
    \footnotesize
    \scriptsize
    \begin{aligned}\label{History_const}
    & \text{X : } y_i: \text{A variable that equals 1 if provider $i$ meets the constraints, and 0 otherwise.}\\
    & \text{D : } \text{The domain of } y_i\text{ is binary, i.e.,} {0, 1}.\\
    & \text{C : } \text{Location constraint ($c_L$): }\forall S \in P.\mathcal{H}: c_L = 1 \Rightarrow S.l = L \text{, where $L$ is a specific microcell.}\\
    & \phantom{\text{C : }} \text{Time constraint ($c_T$): }\forall S \in P.\mathcal{H}: c_T = 1 \Rightarrow S.d = D \text{, where $D$ is a specific time interval.}\\
    & \phantom{\text{C : }} \text{Energy constraint ($c_E$): } \forall S \in P.\mathcal{H}: c_S = 1 \Rightarrow S.a >=  A \text{, where $A$ is a specific energy service size.}\\
    \end{aligned} 
    \end{equation}
    The goal is to find an assignment for the variable $y_i$ that satisfies all the constraints, effectively identifying the providers' history that meets the super-provider's specific constraints. Our CSP formulation is designed to be highly flexible and adaptable to various super-provider preferences. By allowing the super-provider to select any combination of constraints,  we provide a tailored trust assessment that meets their specific requirements. \looseness=-1
    \begin{figure}[!t]
    \centering
     \setlength{\abovecaptionskip}{2pt}
     \setlength{\belowcaptionskip}{0pt}
   \includegraphics[width=0.6\linewidth]{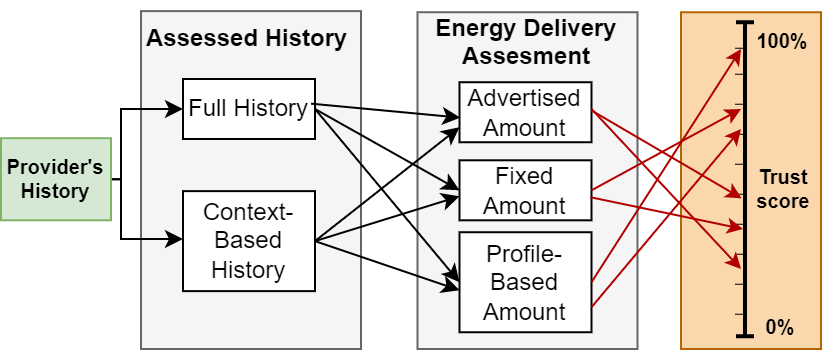}
    \caption{Context-aware trust assessment model}
    \label{fig:TrustAssesment}
  \end{figure}
 
    \item \textbf{Trust Assessment Constraints:} The super-provider's trust constraints are related to the trust assessment attributes and the overall trust score. These constraints are formulated as a tuple of $<W, \alpha, ae>$ where:
    \begin{itemize}[noitemsep,nosep,leftmargin=8pt,labelsep=1pt,itemindent=0pt, labelwidth=*]
    \item $w$ represents the weight assigned to each trust attribute based on its importance, with the constraint that $\sum w_i= 1$  to normalize the impact of each factor on the overall trustworthiness score,
    
    \item $\alpha$ is a threshold for the provider trust score. $\alpha$ is determined based on the super-provider preference. The super-provider might adopt a greedy approach, aiming to maximize energy without considering the providers' trustworthiness. Conversely, they could be risk-averse, prioritizing trust over the amount of provided energy. Alternatively, the super-provider may take a neutral stance, striking a balance between trustworthiness and the energy provided,
    
    \item $ae$ is the adjusted expectations of a super-provider. Super-providers may have different expectations for assessing provider trustworthiness based on their microcell needs. For instance, a provider with a low trust score may be considered trustworthy by a super-provider if they need 50 units, and the provider has historically delivered that amount. Super-providers can determine their expectations according to their requirements, with various parameters such as delivery time and failure impact. \textcolor{black}{In this work, we focus on adjusting expectations concerning energy.  Consequently, we propose three methods for determining the super-provider expectations in terms of energy (see Fig. \ref{fig:TrustAssesment}). This allows the super-provider to choose what best fits their needs:}


    \begin{itemize}[noitemsep,nosep,leftmargin=12pt,labelsep=2pt,itemindent=2pt]
\item Advertised amount: In this setting, a super-provider will assess providers based on the amount of delivered energy against what they advertised using Eq. \ref{eq:ds},\looseness=-1

\item Capped amount: In this setting, a super-provider will assess providers based on the amount of delivered energy against what the super-provider needs ($expected$ $Amount$). For instance, assessing them on delivering 30 units regardless of their service advertisement. The needed amount will be fixed for all providers and may depend on the energy demand of the microcell, the available providers, and their trust score. For instance, if a super-provider needs 100 mAh and has five providers, and their trust score is low, then a possible solution is to assess them by looking into their history and the rate of them delivering 20 units,\looseness=-1

\item Customized amount: In this setting, the super-provider assesses providers using an energy size extracted from their pattern of delivered energy from their history. The reason for adjusting the energy is that sometimes IoT users may overestimate what they can deliver \cite{lakhdari2020Elastic}. The customized amount may be computed using statistical values such as mean, mode, and median, and it will be unique to each provider based on their profile.

\end{itemize}
\textcolor{black}{If the super-provider wants to assess the providers on delivering a fixed amount or a profile-based amount as $expectedAmount$ regardless of their service advertisement. In such a scenario, the denominator in Eq. \ref{eq:ds} will be $expectedAmount$, and the $\mathcal{DS}_{P}$ score will be one if what is provided is larger than $expectedAmount$}.

     \end{itemize}
     \vspace{5pt}
\end{itemize}

\noindent\textcolor{black}{Recall that the context model is used in the context-aware trust assessment phase of the framework (see Fig. \ref{fig:frameWork}). This phase consists of two steps: history filtering and trust assessment. In the following subsections, we will discuss each step of the phase in detail.}
\vspace{-18pt}
\subsubsection{History Filtering:} As previously mentioned, the first step of the trust assessment involves filtering the providers' history.  We employ the history constraints of the context model as a filter to refine the providers' history. \textcolor{black}{The selection of which constraints to apply and the values of the variables will be determined based on the business context and requirements of the super-provider. Recall that the selection of which constraints to apply and the values of the variables in these constraints will be provided as input to our framework and are not the focus of our work.} If a provider lacks sufficient history after applying the constraints, we can gather their history from microcells with similar contexts to our microcell. However, this approach is beyond the scope of this paper. In cases where there is insufficient history (i.e., the number of historical records is less than a predefined threshold), we use the provider's original history for assessment.\looseness=-1
\vspace{-10pt}
\subsubsection{Trust Assessment:} 
The next step uses the proposed trust model to assess the trust of the providers after refining their history in the previous step. \textcolor{black}{The input for this step is the trust constraints from the context model. Note that the super-provider determines any combination of history and energy constraints} (see  Fig. \ref{fig:TrustAssesment}). For example, a super-provider may require to use the full history and a fixed amount assessment. The output is each provider's trust score $\mathcal{P}{trust}$, calculated using Eq. \ref{eq:PTrust}, and the weights of the equation are defined in the context model $W$. $\mathcal{P}{trust}$ is then employed to identify the most trustworthy providers exceeding the super-provider's trust score threshold $\alpha$.\looseness=-1


\vspace{-10pt}
\subsection{Trust-Based Energy Services Composition}
\vspace{-5pt}
This phase aims to compose the most trustworthy energy services to ensure QoE. Given a set of providers and their trust score, we may utilize any priority-based spatio-temporal composition \cite{lakhdari2020composing}\cite{abusafia2022maximizing}\cite{abusafia2020incentive} or other resource allocation algorithms for service provisioning such as Max-Min, knapsack, and genetic algorithms. However, due to the scarcity of energy, a super-provider may end up having a pool of providers with lower trust scores. \textcolor{black}{Intuitively, a super-provider may allocate extra providers as a backup. Hence, we propose a trust-priority heuristic approach that selects additional services to ensure sufficient energy services. Our approach reduces providers' services based on their trust scores and identifies complementary services as backups if the original providers do not fulfill their commitments. In other words, our approach will downsize the providers' energy service size based on their trust score (See Algorithm \ref{alg:TBC}, Lines 5-6).  There are other possible ways to over-provision by increasing the amount of energy demand based on the available providers and their trust score. We leave it for future work to estimate the optimal increase in energy demand using other techniques. Moreover, Moreover, our proposed approach operates on a 'best effort' basis. That is, if the demand is larger than the supply, it will compose the best available set of providers.
Lastly, the super-provider assesses the QoE of the resulting composition using the model discussed in Sec. \ref{prblemform}. The assessment of QoE gives an indicator of consumers' satisfaction in the microcell.}\looseness=-1


\vspace{-10pt}
\section{Evaluation}   \label{ExpSection}
\vspace{-10pt} 
  We investigate the effectiveness and efficiency of the proposed composition approaches based on a comprehensive set of experiments. This section presents a description of the dataset used in the experiments, followed by a  description of the experiment setup and a discussion of the findings.\looseness=-1

\vspace{-10pt} 
\subsection{Dataset Description}
\vspace{-8pt}
We used a real dataset generated from an app developed in \cite{yang2023Monitoring}\cite{yang2022towards}. The app monitors the wireless energy-sharing process that occurs by using coils connected to two smartphones. The consumer determines the granularity of the monitoring time. The app allows users to request energy from nearby smartphones by size, e.g.,1000 mAh, or by time, e.g., to charge for 5 minutes. The dataset consists of energy transfer records between a provider (smartphone) and a consumer (smartphone). The records include attributes such as provider ID, consumer ID, transaction date, time, energy service amount, request amount, and transfer duration. We used the energy dataset to generate QoS parameters for the energy services and requests. For example, the amount of wireless charging transfer in mAh defines the amount of requested/advertised/provided energy. Additionally, the energy dataset records of the wireless charging transfer duration were used to define the end time of each~request/service.\looseness=-1 

We augmented the energy sharing dataset to mimic the behavior of the crowd within microcells by leveraging a dataset published by IBM for a coffee shop chain with three branches in New York City\footnote{https://ibm.co/2O7IvxJ}. The dataset consists of transaction records of customer purchases in each coffee shop over one month. On average, each coffee shop has 560 transactional records per day and 16,500 transaction records in total. We used the IBM dataset to simulate the spatio-temporal features of energy services and requests. Our experiment employs the consumer ID, transaction date, time, location, and coffee shop ID from each record in the dataset to define the spatio-temporal features of energy services and requests, e.g., the start and location of an energy service or a request. We randomly generate service cancellations to create untrustworthy providers.  Table \ref{ExpVariables} presents the experiment parameters and statistics.\looseness=-1

 \vspace{-15pt}
\subsection{Evaluation of the Composition Framework}
\vspace{-5pt}
We compare the proposed heuristic-based composition approached with a baseline traditional resource allocation algorithms, namely, first come first served allocation (\textit{Greedy}), and a priority-based allocation algorithm (\textit{Priority-based}) \cite{kruse2007data}. In Greedy, services are processed based on their start time regardless of their trust score. In Priority-based, services are processed based on the trust score 
We also compared our approach with the knapsack-inspired service composition method (\textit{knapsack-based}) proposed by \cite{lakhdari2020composing}. The approach selects services that maximize the minimum trust value of the participating providers. We also attempted to implement a brute force approach; nonetheless, its substantial computational and memory requirements made effective execution impossible.
\textcolor{black}{We conducted an ablation analysis to assess the impact of different factors, including QoE with different: history constraints, energy delivery assessment constraints, and allocation strategies in different environments. We also examined the cost of incentives and compared the execution time for each method. As mentioned in Sec. \ref{PTrustSec}, multiple methods exist for calculating the weight of the trust assessment attributes, depending on the preference of the super-provider~\cite{bahutair2021multi}.  We assigned all $w$ in Eq. \ref{eq:PTrust} to 0.2 for equal impact on the trust score.}
We ran the approaches in different settings by fixing the energy demand and gradually changing the number of services over the time interval $T$. We repeated the experiment 1000 times at each point and considered the average value for each approach.\looseness=-1
\vspace{-5pt}
\subsubsection{Quality of Experience Evaluation}
\vspace{-5pt}
The first experiment evaluates the impact of the context model's history constraints on QoE. As mentioned earlier, QoE reflects consumer satisfaction over time, and a high QoE for a composition indicates a greater degree of consumer satisfaction. Fig. \ref{fig:QoE_history} displays the average QoE using the \textit{knapsack-based} approach with different constraints, i.e., full history, time-constraints representing the history filtered based on the energy demand's time interval, and spatio-temporal constraints assessing trust using the history at the time interval and the given microcell location.  We set the energy demand in this experiment to 1000 mAh. Fig. \ref{fig:QoE_history} shows that using providers' full history yielded the lowest QoE compared to the context-aware filtered history. This is because assessing trust based on the full history results in many low-trust providers not being used. However, the context-aware trust assessment focuses on their behavior in the microcell. This leads to better assessment and thereby provides more services to allocate, which ensures a higher QoE.\looseness=-1
\begin{table}[!t]
\noindent
    \begin{minipage}{\linewidth}
      \begin{minipage}[b]{0.43\linewidth}
        \centering
        
        \caption{Experiments Variables}
        \label{ExpVariables}
        \resizebox{\linewidth}{!}{%
        \renewcommand{\arraystretch}{1}
        \begin{tabular}{|l|l|}
        \hline
      Variables & Values   \\ \hline
        Total energy service records&  \\ 
        for all coffee shops             & 49894   \\ 
        Energy providers                  &{2248}              \\ 
        Duration of all energy services             & {10 - 30 minutes} \\ 
        Energy services amount                  & {150 - 300 mAh}      \\
        Energy  demand amount                 & {500 - 2500 mAh}       \\
        Time window             & {2 hours} \\ \hline
        \end{tabular}}%
          \end{minipage}
      \hfill
     \begin{minipage}{.5\linewidth}
        \centering
       
         \setlength{\abovecaptionskip}{0pt}
             \setlength{\belowcaptionskip}{-1pt}
       \includegraphics[width=0.9\linewidth]{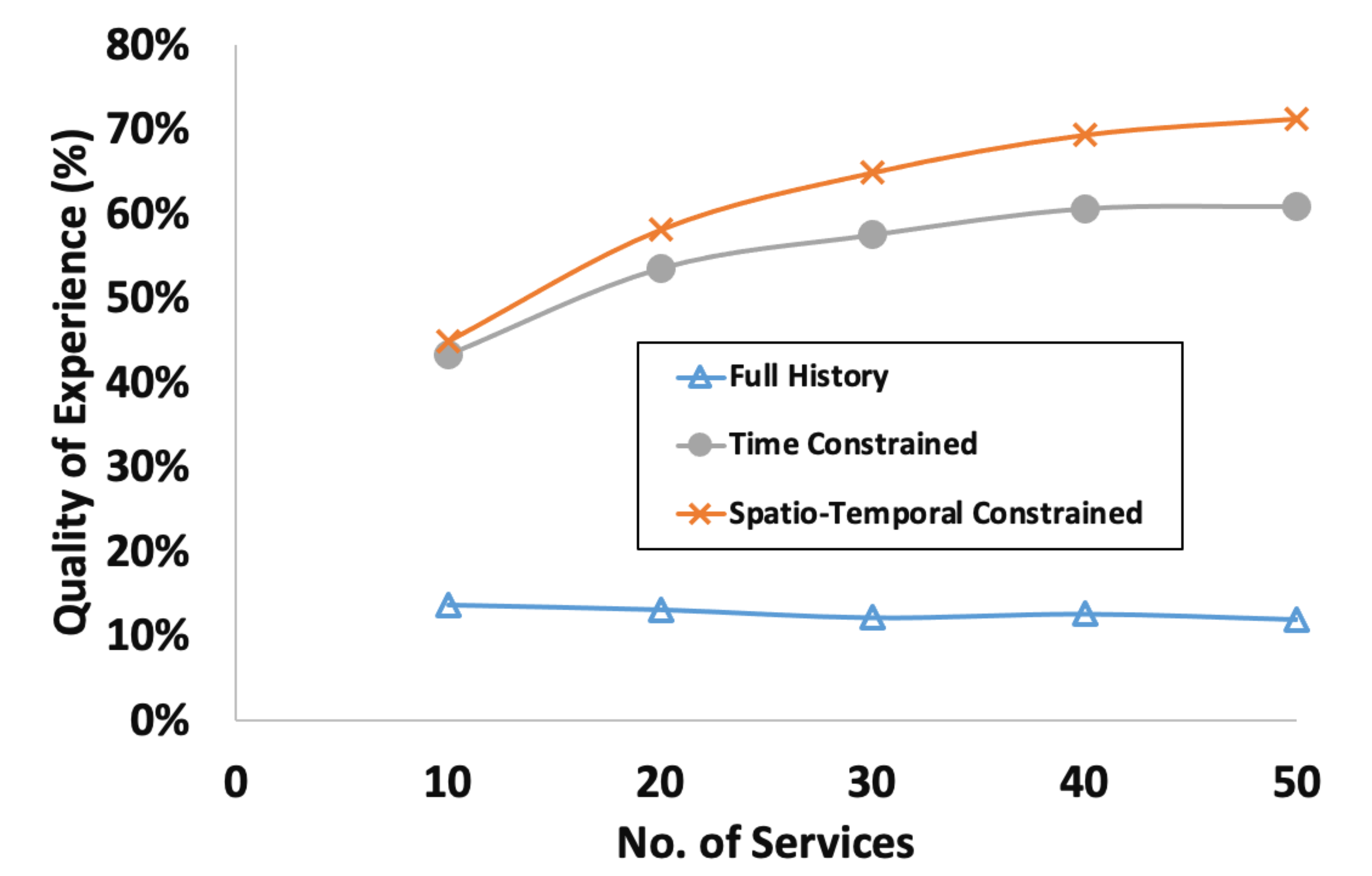}
    \captionof{figure}{The average of QoE using different history }
    \label{fig:QoE_history}
        \end{minipage}
            \end{minipage}
\end{table}
 \begin{figure}[!t]
\centering
\begin{minipage}{.48\textwidth}
\centering
 \setlength{\abovecaptionskip}{0pt}
\setlength{\belowcaptionskip}{-1pt}
\includegraphics[width=0.9\linewidth]{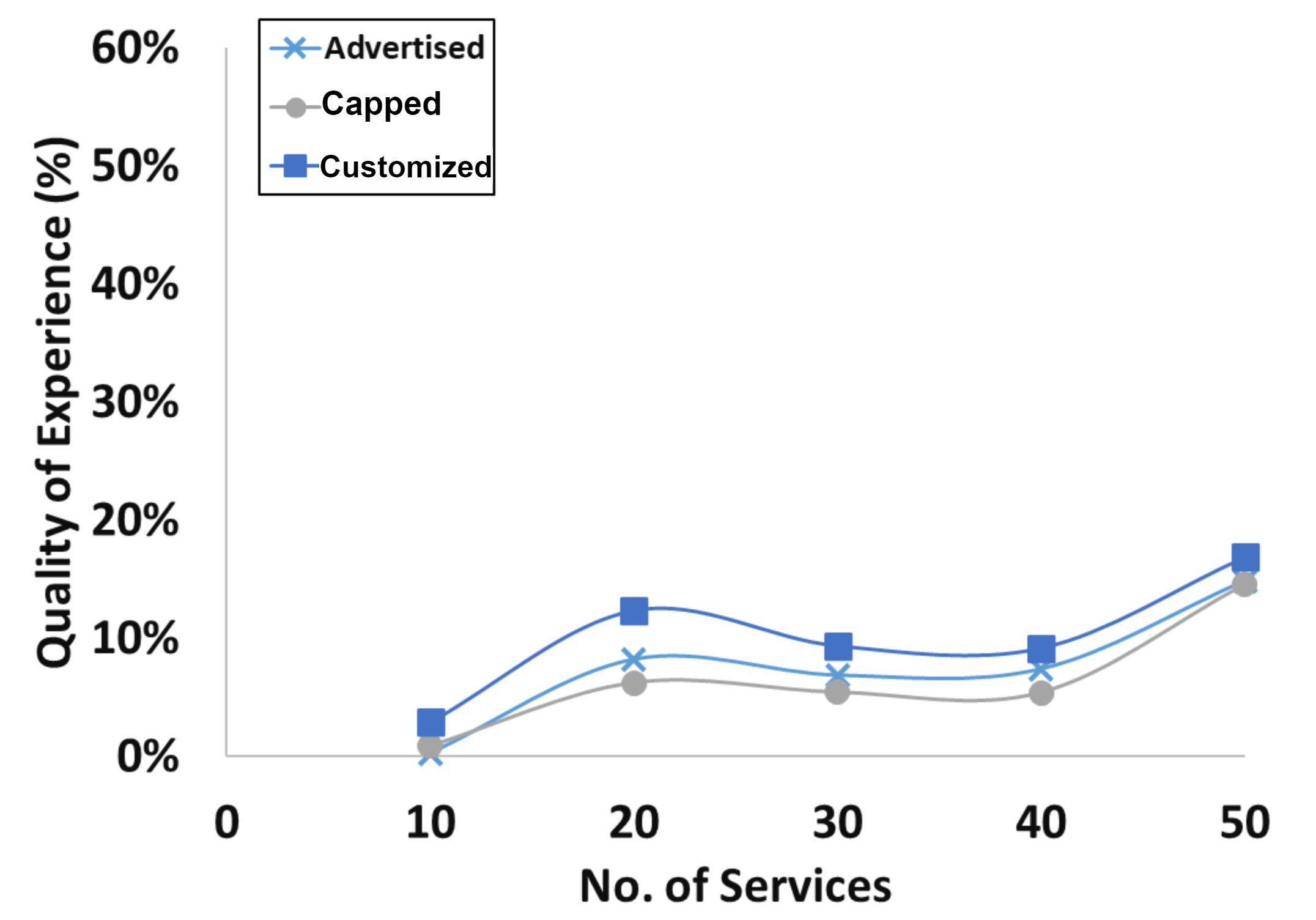}
\caption{The average of QoE using different trust assessments}
\label{fig:QoE_Adjusted}
\end{minipage}
\hfill
\begin{minipage}{.48\textwidth}
\centering
 \setlength{\abovecaptionskip}{0pt}
     \setlength{\belowcaptionskip}{-1pt}
\includegraphics[width=1\linewidth]{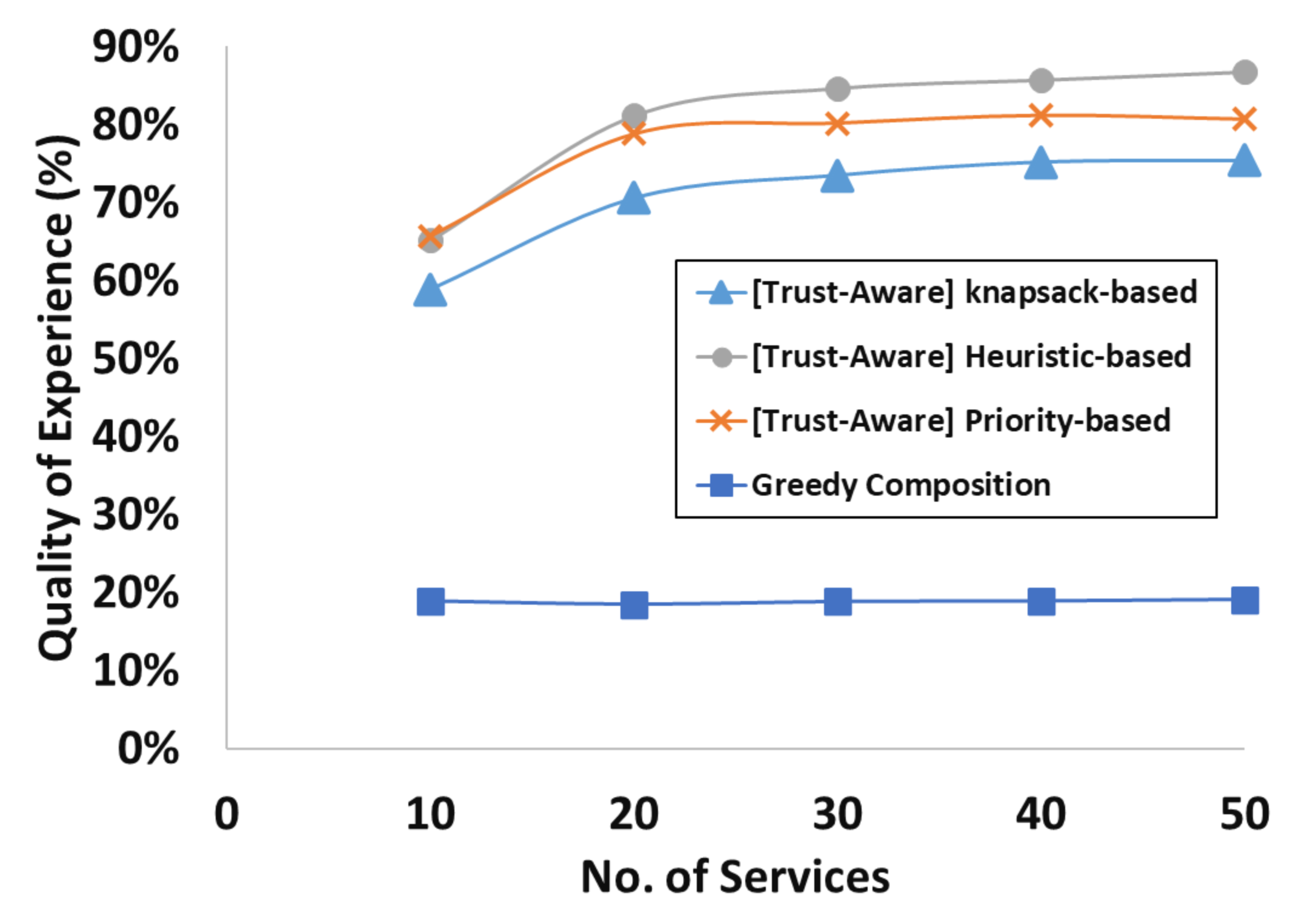}
\caption{The average of QoE in trustworthy environment}
\label{fig:QoE_Trust}
\end{minipage}      
\end{figure}

\begin{figure}[!t]
\centering
\begin{minipage}{.48\textwidth}
\centering
 \setlength{\abovecaptionskip}{0pt}
\setlength{\belowcaptionskip}{0pt}
\includegraphics[width=\linewidth]{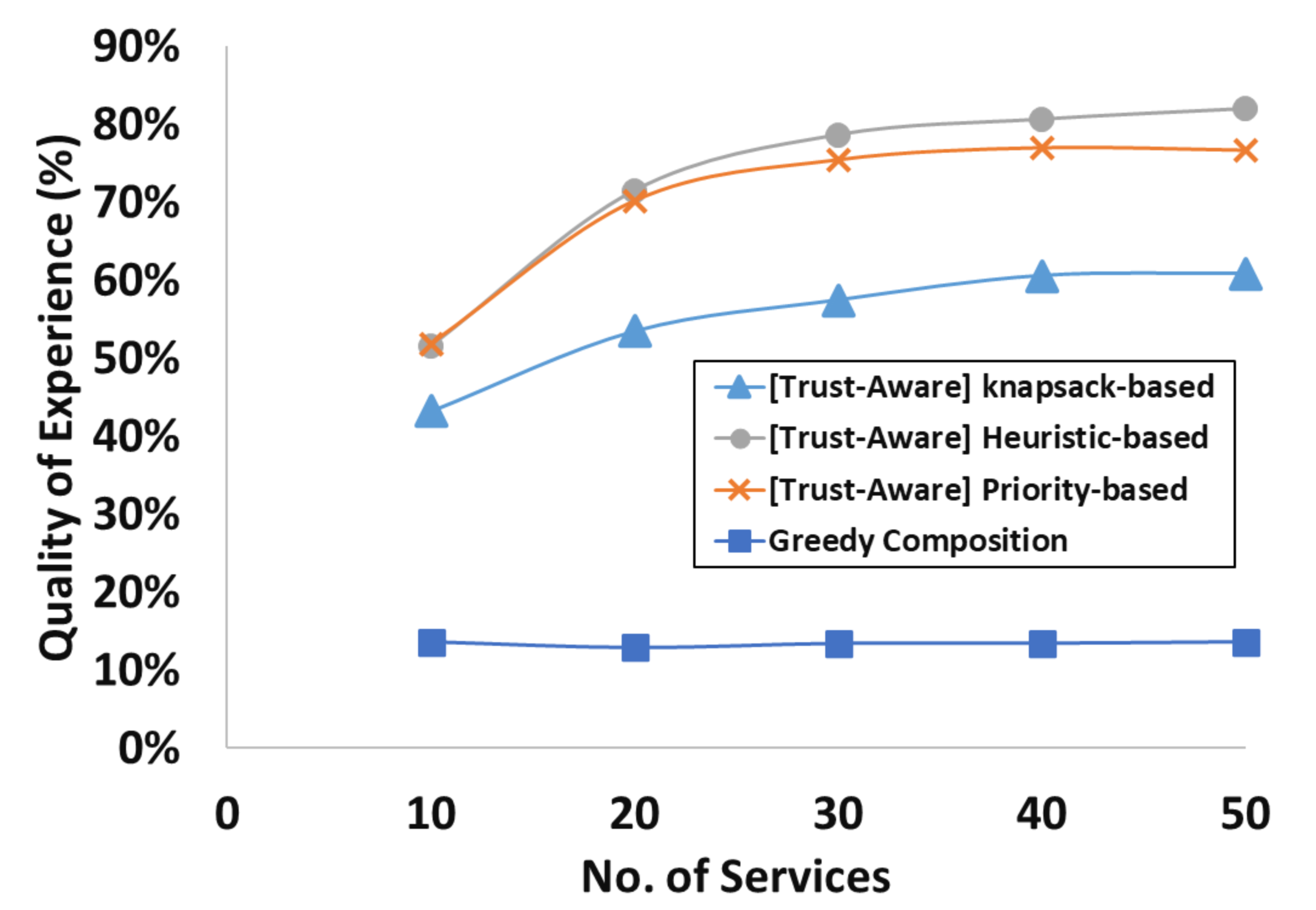}
\caption{The average of QoE in neutral environment}
\label{fig:QoE_neutral}
\end{minipage}
\hfill
\begin{minipage}{.48\textwidth}
\centering
 \setlength{\abovecaptionskip}{0pt}
     \setlength{\belowcaptionskip}{0pt}
\includegraphics[width=1\linewidth]{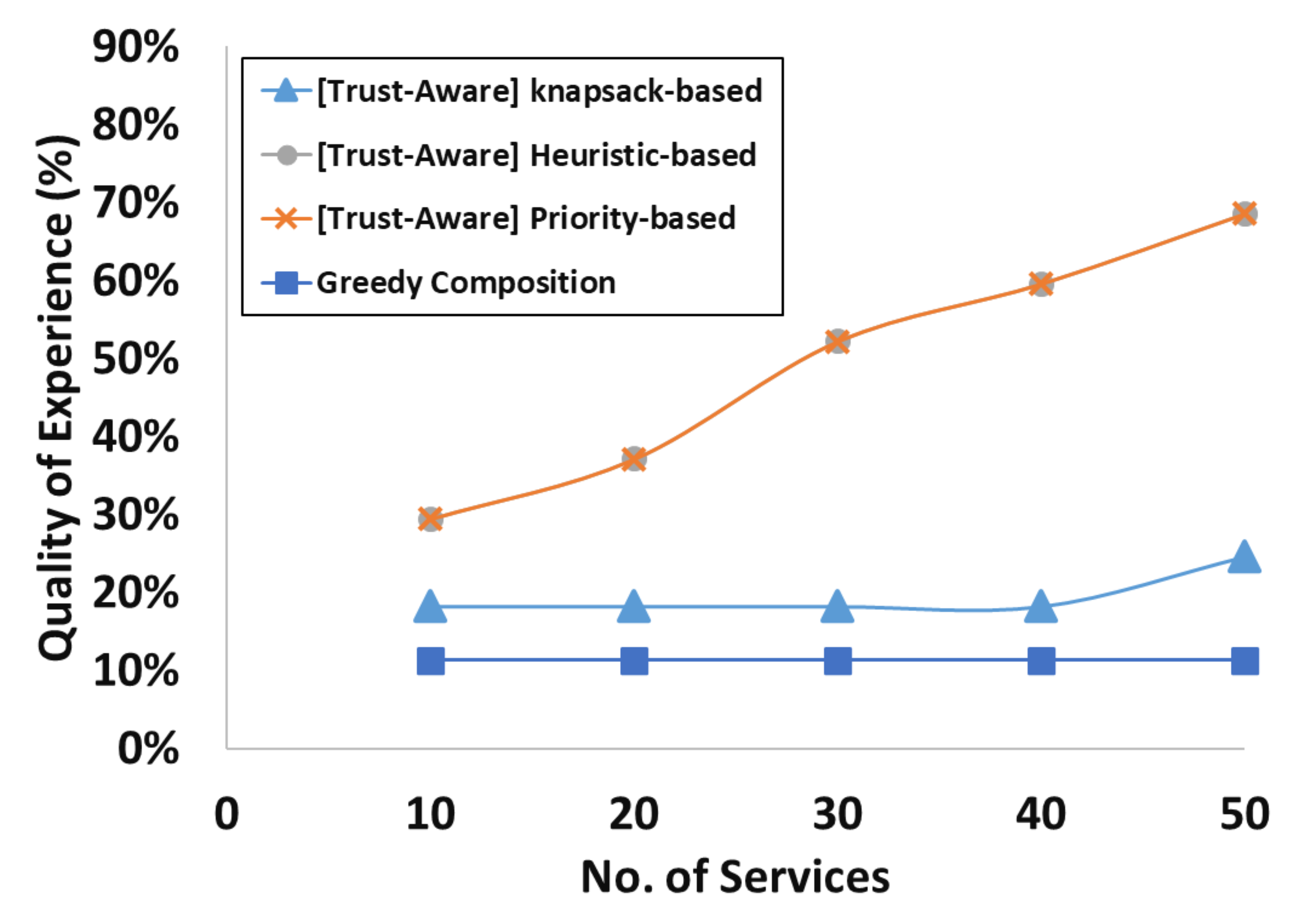}
\caption{The average of QoE in untrustworthy environment}
\label{fig:QoE_untrustworthy}
\end{minipage}      
\end{figure}

The second experiment examines the impact of the context model's trust assessment constraints on QoE. Fig. \ref{fig:QoE_Adjusted} displays the average QoE using the \textit{knapsack-based} approach with different expected energy assessments: "advertised" assesses providers based on the service amount advertised in their history record, "capped" assesses them based on a fixed amount determined by the super-provider (e.g., 50 units), and "customized" is a profile-based assessment. In the customized assessment, we evaluated providers based on the median of their historical records. Fig. \ref{fig:QoE_Adjusted} shows a slight QoE improvement when assessing providers based on the median; further experiments are needed to understand the consistency in providers' patterns and the sufficient number of records for profiling a user. We plan to explore the provider's patterns in the future.\looseness=-1

\begin{figure}[!t]
\centering
\begin{minipage}{.48\textwidth}
\centering
 \setlength{\abovecaptionskip}{0pt}
\setlength{\belowcaptionskip}{0pt}
\includegraphics[width=\linewidth]{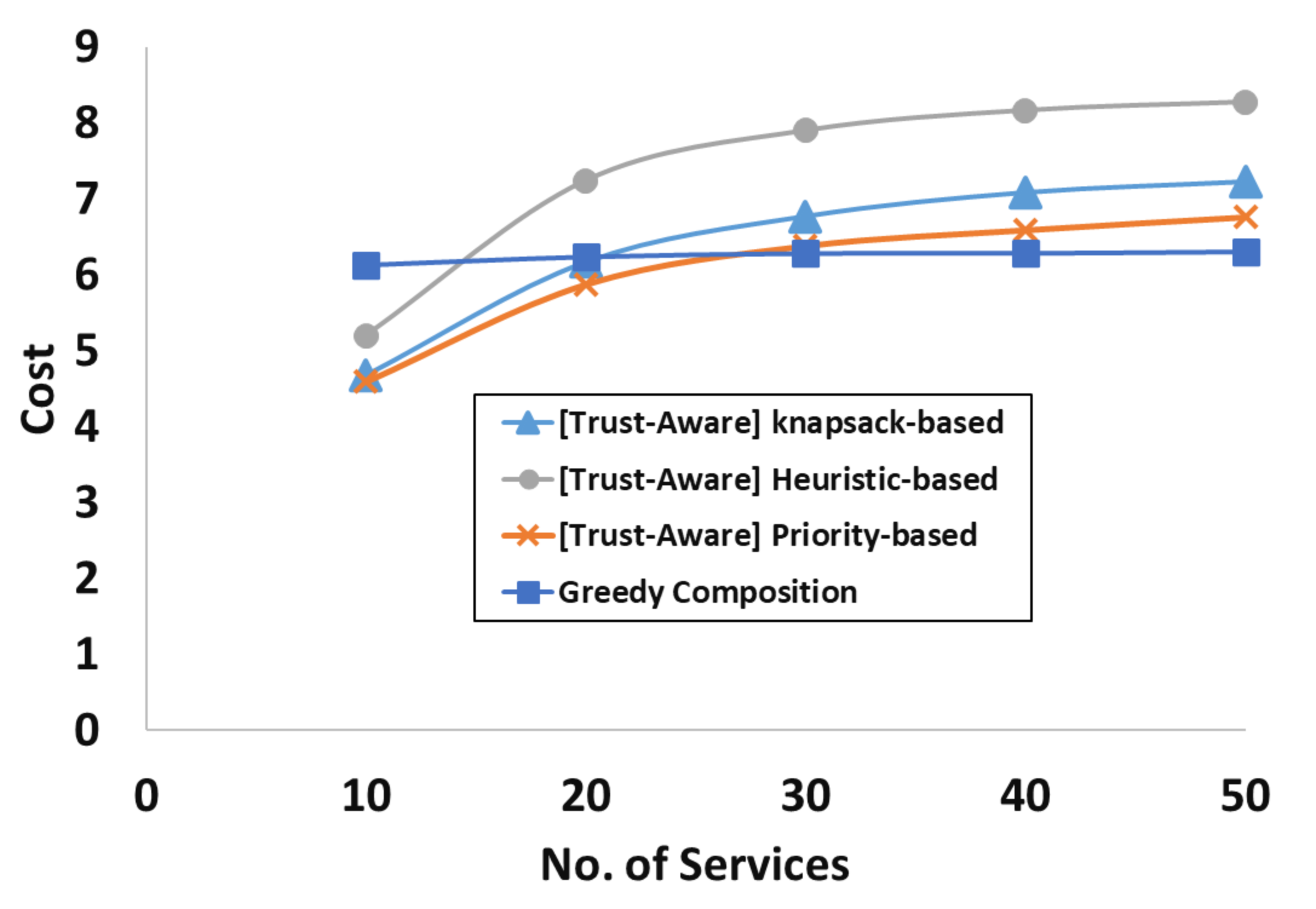}
\caption{The average of cost}
\label{fig:R_cost}
\end{minipage}
\hfill
\begin{minipage}{.48\textwidth}
\centering
 \setlength{\abovecaptionskip}{0pt}
     \setlength{\belowcaptionskip}{0pt}
\includegraphics[width=1\linewidth]{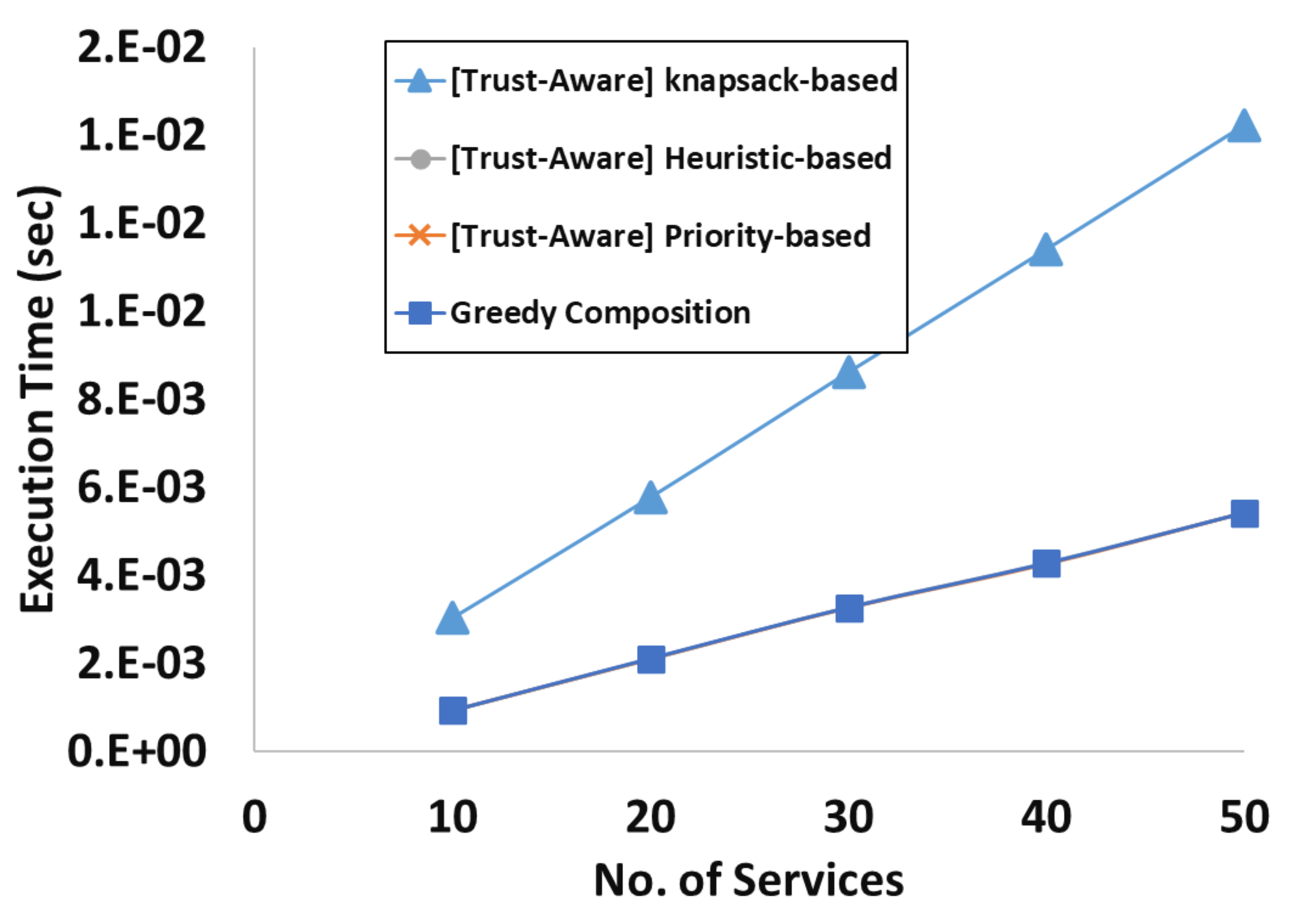}
\caption{The average of computation cost}
\label{fig:R_exe}
\end{minipage}      
\end{figure}

The third experiment evaluates the effectiveness of our proposed heuristic-based composition. We compare our approach with the aforementioned approaches in three environment settings: a trustworthy environment where most of the provider's trust score is high, i.e., above 80\% (see Fig. \ref{fig:QoE_Trust}); a neutral environment where providers' trustworthiness follows a random distribution (see Fig. \ref{fig:QoE_neutral}); and an untrustworthy environment where most of the provider's trust score is low, i.e., below 20\% (see Fig. \ref{fig:QoE_untrustworthy}). Overall, our proposed approach performs better than the rest of the approaches due to its over-provisioning strategy. However, the over-provisioning comes with a higher cost of rewards (See Fig. \ref{fig:R_cost}). We compute the cost as the price per energy unit multiplied by the amount of provided energy \cite{abusafia2020incentive}. Another observation is that QoE increases for all approaches except greedy in all settings. This is intuitive, as with the increase in services, there is more energy to fulfill the energy demand and thereby increase QoE.\looseness=-1

\vspace{-8pt}
\subsubsection{Computation Efficiency Evaluation}
\vspace{-8pt}
In the fourth experiment, we assessed the computational cost of all approaches. The execution time for all approaches increases with the increase in services' availability (see  Fig. \ref{fig:R_exe}). This is due to the increase in processing time to assign these services.\looseness=-1

\vspace{-15pt}
\section{Related Work}
\vspace{-10pt}
The background of our work comes from energy services and trust assessment in IoT services. We present the related work to our research in each domain.\looseness=-1
\vspace{-15pt}
\subsubsection*{Crowdsourcing Energy Services:}
Energy exchange services have emerged as alternative solutions for charging IoT devices \cite{lakhdari2020Vision}\cite{dhungana2020peer}. Several studies have addressed challenges related to meeting the demands of energy consumers \cite{lakhdari2020composing}\cite{lakhdari2020Elastic}\cite{lakhdari2020fluid}. A time-based composition algorithm was introduced to compose energy services to satisfy consumers’ energy needs \cite{lakhdari2020composing}. The algorithm suggests using partial services and a fractional knapsack to maximize the provided energy.  The intermittent nature of energy services has been addressed using a fluid approach \cite{lakhdari2020fluid}.  Other research has addressed challenges from the provider's perspective \cite{abusafia2020incentive}\cite{abusafia2022services}. A context-aware incentive model was suggested to address the resistance to offering energy services \cite{abusafia2020incentive}. Another study proposed a model to estimate the energy loss in sharing energy services \cite{yang2023energy}. Recent studies have addressed these challenges from a super-provider's perspective \cite{abusafia2023flow}\cite{abusafia2022maximizing} \cite{abusafia2022quality}. A QoE model was suggested as a key indicator for allocating services to requests \cite{abusafia2022maximizing} \cite{abusafia2022quality}. \textcolor{black}{Neither QoE-based composition approach considers providers' uncertain availability.  To the best of our knowledge, challenges related to the uncertainty of providers' provisioning remain unaddressed.}\looseness=-1

\vspace{-15pt}
\subsubsection*{Trust assessment in IoT services:} 
Trust assessment in crowdsourced IoT service environments is fairly new. Most of the proposed trust approaches rely on either previous experiences \cite{kantarci2014mobility} or social relations \cite{Cao2015} to assess trust. The work in \cite{kantarci2014mobility} proposed a framework that assesses service providers based on their reputations by a central authority. A framework was proposed in \cite{Cao2015} to eliminate the privacy risks associated with public Wi-Fi hotspots. The proposed framework lacks generality as it can only be used for Wi-Fi hotspot services. 
\textcolor{black}{Another study proposed a QoE-based trustable framework for managing services in mobile edge computing (MEC) \cite{tahaei2018qoe}. However, its focus on MEC system layers and lack of clear trust definition or experimental validation limits its applicability in our context. Another study proposes a trusted resource allocation mechanism for fog computing where a fog-to-fog offloading scheme balances the load on the fog layer \cite{jain2022trusted}. However, the approaches mentioned above may not be ideal for IoT crowdsourcing environments. Trust assessment in such environments presents unique challenges due to characteristics like the diversity and anonymity of IoT users and devices and the lack of a central managing authority \cite{bahutair2021multi}. These characteristics make it more difficult to acquire an accurate measure of trust.  A recent study proposed to assess trust in crowdsourcing IoT environments from multiple perspectives, such as device, owner, and service \cite{bahutair2021multi}. Other solutions proposed to assess trust from a usage-based perspective \cite{ba2022multi}.  A trust assessment method is proposed to evaluate services based on their usage. For example, a user streaming a video may have lower trust requirements than another user engaged in online banking activities \cite{ba2022multi}. 
These frameworks are not applicable in IoT energy environments due to the energy scarcity, fluctuating behavior of IoT users, and influence of super-provider expectations on trust scores \cite{lakhdari2020fluid}.} To the best of our knowledge, no solution assesses the trustworthiness of providers' provisioning while considering the super-providers' expectations and requirements.\looseness=-1
\vspace{-5pt}
 \section{Conclusion}
\vspace{-8pt}
\textcolor{black}{The allocation of energy services has been proposed as a tool to assure consumers' Quality of Experience (QoE). However, the existing frameworks assume that providers will always deliver the advertised service. In contrast, the dynamic nature of the energy services environment may result in uncertainty in providers' commitment. Therefore, a trust assessment is required to evaluate the trustworthiness of providers. Existing trust assessments are not applicable due to the energy scarcity and the dynamic behavior of IoT users.} Consequently, this paper proposes a novel context-aware trust assessment model. The proposed model assesses providers' trustworthiness in various contexts, considering their behavior and energy provisioning history. Moreover, a trust-adaptive composition technique is proposed for optimal energy allocation, ensuring efficient energy service provisioning. We conducted a set of experiments to assess the performance of the proposed framework. Experiments showed that filtering history based on the microcell context and adjusting expectations based on the super-provider's needs resulted in better trust assessment and improved QoE. The future direction is to consider the probability of change in the energy demand.\looseness=-1
\vspace{-10pt}
\section*{Acknowledgment}
This research was partly made possible by LE220100078 and DP220101823 grants from the Australian Research Council. The statements made herein are solely the responsibility of the authors.

\vspace{-15pt}
\bibliographystyle{unsrt}
\bibliography{main}

\end{document}